\documentclass{article}

\usepackage{spconf2,amsmath,amssymb,amsfonts,graphicx,hyperref}
    
\usepackage{bm} 

\usepackage{cite}
\usepackage{textcomp}
\usepackage{xcolor}
\def\BibTeX{{\rm B\kern-.05em{\sc i\kern-.025em b}\kern-.08em
    T\kern-.1667em\lower.7ex\hbox{E}\kern-.125emX}}

\usepackage[ruled, lined, linesnumbered, commentsnumbered, longend]{algorithm2e}
\usepackage{algpseudocode}

\usepackage{caption}
\usepackage{url}

\usepackage[top=0.68in, bottom=0.68in, left=0.6in, right=0.6in]{geometry}  

\newcommand{\VEC}[1]{\boldsymbol{#1}}
\newcommand{\MAT}[1]{\boldsymbol{#1}}
\newcommand{\FIELD}[1]{\mathbb{#1}}
\newcommand{\PR}[1]{\left ( {#1} \right ) }
\newcommand{\BK}[1]{\left [ {#1} \right ] }
\newcommand{\CR}[1]{\left \{ {#1} \right \} }
\newcommand{\FLOOR}[1]{\left \lfloor {#1} \right \rfloor}

\newcommand{\ABS}[1]{\left | {#1} \right |}

\newcommand{\NORM}[1]{\left\lVert#1\right\rVert}
\newcommand{\ARGMIN}[1]{ \arg\min_{#1}}

\newcommand{\DERIVDERIVFLAT}[2]{ {d^2 {#1} }/{d  {#2}^2 } }
\newcommand{\DERIVFLAT}[2]{ {d  {#1} }/{d  {#2}} }

\newcommand{\PDERIV}[2]{ \frac{\partial {#1} }{\partial  {#2}} }
\newcommand{\PDERIVFLAT}[2]{ {\partial  {#1} }/{\partial  {#2}} }
\newcommand{\PDERIVTWO}[3]{ \frac{\partial^2 {#1} }{\partial {#2} \, \partial  {#3} } }

\newcommand{\DIAGF}[1]{\textbf{\textrm{diag}} \PR{#1} }

\newcommand{\SGN}[1]{\textbf{\textrm{sgn}} \left ( {#1}  \right ) }

\newcommand{\TOEPLITZBND}[3]{\boldsymbol{T}_{#2}^{#3}\left ( {#1} \right ) }

\newcommand{\VECT}[1]{\textrm{vec}\PR{#1} }

\DeclareMathSymbol{\minus}{\mathbin}{AMSa}{"39}

\SetKwInput{KwRequire}{Require}

\title{Constraint Optimized Multichannel Mixer-limiter Design}

\name{1\textsuperscript{st} Yuancheng Luo, $\qquad$ 2\textsuperscript{nd} Dmitriy Yamkovoy, $\qquad$ 3\textsuperscript{rd} Guillermo Garcia}
\address{Amazon.com, \emph{\{mikeluo, dpy, guigarg\}@amazon.com} }

\begin{document}

\ninept

\maketitle

\begin{abstract}
Multichannel audio mixer and limiter designs are conventionally decoupled for content reproduction over loudspeaker arrays due to high computational complexity and run-time costs. We propose a coupled mixer-limiter-envelope design formulated as an efficient linear-constrained quadratic program that minimizes a distortion objective over multichannel gain variables subject to sample mixture constraints. Novel methods for asymmetric constant overlap-add window optimization, objective function approximation, variable and constraint reduction are presented. Experiments demonstrate distortion reduction of the coupled design, and computational trade-offs required for efficient real-time processing.
\end{abstract}


\setlength{\belowdisplayskip}{6pt}
\setlength{\belowdisplayshortskip}{6pt}
\setlength{\abovedisplayskip}{6pt}
\setlength{\abovedisplayshortskip}{6pt}

\setlength{\belowcaptionskip}{-2.0pt}
\setlength{\intextsep}{7pt}

\setlength{\skip\footins}{1mm}

\vspace{-3.4mm}

\section{Introduction}
\label{SEC:INTRO}

Multichannel audio content reproduction over loudspeaker arrays has grown in popularity in recent years with the proliferation of low-cost sound-bar and smart-speaker consumer electronics. Such audio-reproduction systems are typically resource-constrained compared to professional-grade loudspeakers in terms of available transducer-level digital-electrical headroom, and acoustic output. A conventional method digitally compresses the audio content's dynamic-range throughout channel-mixing stages to stay under digital full-scale levels, maximize loudness, and satisfy studio standards \cite{EBU_2011, COLLOMS_2018, TOOLE_2017, CHIARELLA_2006}. A channel matrix-mixer \cite{ITU_755_4, HERRE_2008} or matrix-decoder \cite{DRESSLER_DOLBY_2000, GRIESINGER_1996, MEARES_1998} therefore allocates headroom between input channels and output transducers;  dynamic-range controllers (DRCs) \cite{MCNALLY_DRC_1984, GIANNOULIS_2012, KATES_2005} such as peak-limiters are thereby placed downstream for transducer and amplifier protection. 

Several deficiencies of the conventional method are known: A multichannel matrix-mixer can conservatively pre-allocate headroom for each input channel such that maximum channel mixture levels minimally activate downstream DRCs. Pre-allocation however distorts the audio mixture in the absence of run-time monitoring; a channel's upper dynamic range may be unnecessarily lowered when other channels are sparse, and the choice of mixing gains can alter the spectral and channel balance of the original content. Terminal DRCs operating on mixtures of channels per transducer can limit at different times, intermittently distorting both channel balance and the loudspeaker array's directivity.

We address both pre-allocation and terminal-limiter problems by coupling time-varying channel-gain reduction with per-sample constraints of the channel mixtures via a sequence of quadratic programming (QP) problems. Section \ref{SEC:QPL} presents our QP mixer-limiter design, and relates the QP's feasibility with a novel constant overlap-add (COLA) \cite{SMITH_JULIUS_2011} constrained gain envelope construction. Section \ref{SEC:OBJ} presents the channel-mixture's distortion objective and derives the QP objective from the former's optimal Taylor series approximation. Section \ref{SEC:VCR} extends the QP formulation to joint multi-band \cite{SCHMIDT_DRC_MULTIBAND_1996} multi-content mixers, and introduces novel variable and constraint size reduction methods for efficient computation. Section \ref{SEC:EXP} shows experimental results for distortion reduction and computational performance.

\section{Quadratic Program Mixer-Limiter}
\label{SEC:QPL}

Let  $\MAT{S} \in \FIELD{R}^{F \times N}$ be a matrix of $N$ input channel column-vectors of $F$ samples in an audio frame. Each of the $N$ input channels are mixed with independent variable gains in vector $\VEC{x} = \BK{x_1, \hdots, x_N}^T$ to produce a single output channel mixture limited in dynamic range and satisfying the equivalent modulus linear constraints given by
\begin{equation}
\begin{split}
& -\tau  \leq \sum_{n=1}^N S_{mn} x_{n} \leq \tau,  \quad \tau \geq 0, \quad  1 \leq m \leq F, \\
\end{split}
\label{EQ:QPL:CONSTR_SAMP}
\end{equation}
where $\tau$ is a user-specified non-negative threshold for any mixture of samples within a frame. 
The variable gains are subject to non-negative bounds under unity and are equivalent to box-constraints given by
\begin{equation}
\begin{split}
 & 0    \leq  l_n \leq x_{n} \leq u_n \leq 1, \quad    1 \leq n \leq N, \\ 
\end{split}
\label{EQ:QPL:CONSTR_GAIN_ABS}
\end{equation}
and therefore only apply gain reduction to each input channel.
The gain variables $\VEC{x}$, constrained to the feasible space that satisfy the linear constraints, are then found by minimization over a quadratic polynomial objective function. The latter's standard form is given by 
\begin{equation}
\begin{split}
f(\VEC{x}) = \frac{1}{2} \VEC{x}^T \MAT{Q} \VEC{x} + \VEC{c}^T \VEC{x} + d,
\end{split}
\label{EQ:QPL:OBJ}
\end{equation}
where symmetric matrix $\MAT{Q} \in \FIELD{R}^{N \times N}$, vector $\VEC{c} \in \FIELD{R}^{N \times 1}$, and constant $d$ parameterize our distortion objective in section \ref{SEC:OBJ}.
Minimizing \eqref{EQ:QPL:OBJ} subject to constraints \eqref{EQ:QPL:CONSTR_SAMP}, \eqref{EQ:QPL:CONSTR_GAIN_ABS} in vectorized form is given by
\begin{equation}
\begin{split}
 \VEC{x}_*   & = \ARGMIN{\VEC{x}} f(\VEC{x}), \quad   \textrm{s.t. } 
-\VEC{\tau} \leq  \MAT{S} \VEC{x} \leq \VEC{\tau},  \quad 
\VEC{l}    \leq \VEC{x} \leq \VEC{u} \leq \VEC{1},
\end{split}
\label{EQ:QPL:CONSTR}
\end{equation}
where $\VEC{\tau} = \tau \VEC{1} = \tau \sum_{i=1}^F \VEC{e}_i \in \FIELD{R}^{F \times 1}$ is the vector of constant threshold $\tau$ over frame-samples, $\VEC{e}_i$ is the standard basis, and  $\VEC{l}, \VEC{u} \in \FIELD{R}^{N \times 1}$ are the vectors of lower and upper bound respectively. We now construct the limiter's gain envelope across frame-wise solutions to \eqref{EQ:QPL:CONSTR}.

\textbf{Constrained Limiter-Envelope Design:}
Attack, hold, and release-time parameters are commonly used to restrict the velocity and shape of a limiter's gain envelope,  constraining the latter to be smooth and reducing any audible distortion when multiplied by the input signals \cite{EARGLE_1996};  attack refers to the early portion of the gain envelope of increasing gain reduction, hold is the middle portion of constant gain reduction, and release is the late portion of decreasing gain reduction. We can define an envelope function with attack, hold, release dynamics, and supports over the solutions to \eqref{EQ:QPL:CONSTR} across overlapped audio frames with look-ahead. Given an audio stream $Y(m, n)$ of $N$ input channels for sample index $m$ and channel index $n$, let $\MAT{S}^{\CR{k}} \in \FIELD{R}^{\bar{F} \times N}$ be the $k^{th}$ audio frame of size $F$ augmented with $L$ look-ahead samples per channel, where $\bar{F} =  F + L$, be defined as follows:
\begin{equation}
\begin{split}
S^{\CR{k}}_{mn} = \left \{ \begin{array}{cc} Y(m + (k-1)F, \, n), & 1 \leq m \leq  F + L \\ 0, & \textrm{otherwise} \end{array}\right . .
\end{split}
\label{EQ:QPL:STREAM}
\end{equation}
For the $k^{th}$ augmented frame, define the QP minimizer $\VEC{x}_{*}^{\CR{k}}$ s.t. the sample constraint set $ \VEC{\xi}^{\CR{k}}$ in the frame and look-ahead as follows:
\begin{equation}
\begin{split}
& \qquad \, \,   \VEC{x}_{*}^{\CR{k}}  = \ARGMIN{\VEC{x}} f(\VEC{x}) \quad \textrm{s.t. linear constraints }  \VEC{\xi}^{\CR{k}}, \\ 
 \VEC{\xi}^{\CR{k}} & = \left \{\begin{array}{rl} -\VEC{\tau} \leq \MAT{S}^{\CR{k}} \VEC{x} \leq \VEC{\tau}, & 2 \bar{F} \textrm{ Mixture limits \eqref{EQ:QPL:CONSTR_SAMP}}\\
   \VEC{0} \leq \VEC{x} \leq \VEC{u} \leq \VEC{1}, & 2N \textrm{ Variable bounds  \eqref{EQ:QPL:CONSTR_GAIN_ABS}} \\ 
   \end{array} \right .,  \\
\end{split}
\label{EQ:QPL:QPFRAME}
\raisetag{1\baselineskip}
\end{equation}
where $\MAT{S}^{\CR{k}}$ replaces $\MAT{S}$ in \eqref{EQ:QPL:CONSTR}, $\VEC{\tau} = \tau \VEC{1} \in \FIELD{R}^{\bar{F} \times 1}$, and the variable lower bounds are set to $\VEC{0}$ to ensure that a feasible solution exists. The enveloped output mixture $y(t)$ at time $t$ is therefore given by the sum of gain enveloped inputs and the envelope function $v: \FIELD{R} \rightarrow \FIELD{R}^{0+}$:
\begin{equation}
\begin{split}
 y(t)  = \sum_{n=1}^N Y(t, n) v_n(t), \quad
 v_n(t)  = \sum_{k=0}^{\infty} W_n(t - kF)  x_{*n}^{\CR{k+1}},
\end{split}
\label{EQ:QPL:ENVELOPE}
\end{equation}
where $v_n(t)$ are weighted mixtures of solutions across frames. We show the bounds $-\tau \leq y(t) \leq \tau$ and $0 \leq  v_n(t)  \leq 1$  are satisfied via the design of a weighting function $W_n(t)$.

\begin{figure}[h]
  \centering
  {\includegraphics[width=0.85\linewidth]{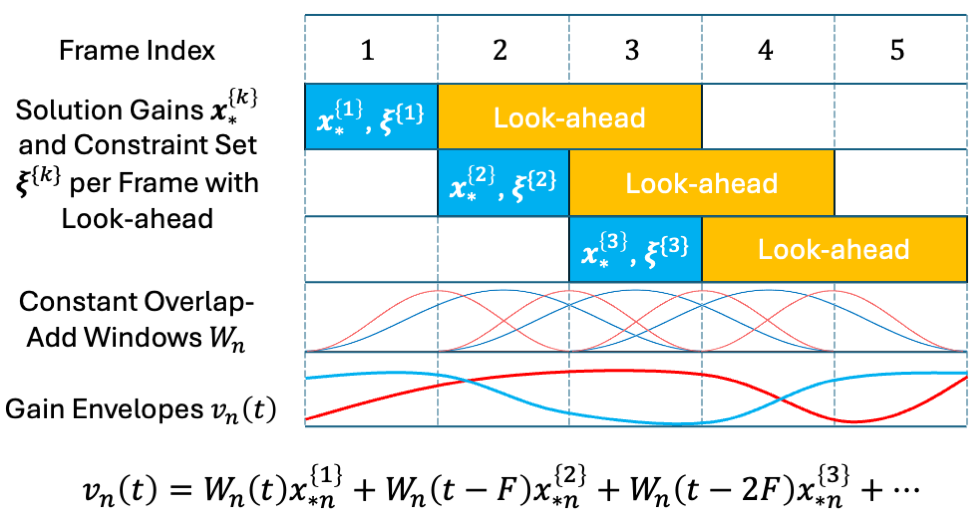}}
\caption{Blue and red envelopes $v_n(t)$ are COLA weighted mixtures of solutions in \eqref{EQ:QPL:ENVELOPE}, belonging to different input channels, and computed over augmented frames $\MAT{S}^{\CR{k}}$ in \eqref{EQ:QPL:STREAM} with different look-ahead times.}
\label{FIG:QPL:ENV}
\end{figure}

Observe in Fig. \ref{FIG:QPL:ENV} that the convex set of solutions from consecutive frames $ \alpha \VEC{x}_*^{\CR{k}}  + (1 - \alpha) \VEC{x}_*^{\CR{k \pm 1}}, \, 0 \leq \alpha \leq 1$ is non-empty for the half-space intersection of $\VEC{\xi}^{\CR{k}} \cap \VEC{\xi}^{\CR{k \pm 1}}$ common mixture-limit constraints from overlapped look-ahead and frame samples of $\MAT{S}^{\CR{k}}, \MAT{S}^{\CR{k \pm 1}}$ respectively. The set of solutions $v_n(t)$ in \eqref{EQ:QPL:ENVELOPE}, evaluated across multiple consecutive frames, is convex and satisfies the mixture-limited constraints of \eqref{EQ:QPL:CONSTR_SAMP} at time $t$ if $W_n(t)$ are window functions with bounded non-negative supports in the frame and look-ahead interval, zero-valued elsewhere, and have the COLA \cite{SMITH_JULIUS_2011} property given by
\begin{equation}
\begin{split}
\left \{ \begin{array}{rc} 0 \leq W_n(t) \leq 1, & 1 \leq t \leq F + L  \\ W_n(t) = 0, & \textrm{otherwise} \end{array} \right \}, & \quad \textrm{Bounded supports} ,\\
    \sum_{k = \minus \infty}^{\infty} W_n(t - kF) = 1, \quad \forall t \in \FIELD{R}, & \quad \textrm{COLA property}.
\end{split}
\label{EQ:QPL:WINDOW}
\end{equation}
For bounded time $t$, the gain envelopes  $\VEC{v}(t) = \BK{v_1(t), \hdots, v_N(t)}^T$ span Barycentric weighted $\VEC{x}_*^{\CR{k}}$ of consecutive sequences of frame index $k$ satisfying $1+(k-1)F \leq t \leq (k-1)F + \bar{F}$. We therefore design window function $W_n(t)$ with bounded COLA and dynamics constraints for smoothly transitioning between supports or frames as to minimize discontinuities in the enveloped output mixture $y(t)$.


\textbf{Dynamics Constrained COLA Window Design:}
We can approximate a smooth COLA window $W(t)$ with characteristic attack, hold, and release dynamics via constrained optimization over the uniformly-sampled and integer-spaced window samples $\omega(t) = W(t), \, t \in \FIELD{Z}$. Let the characteristic dynamics of $W(t)$ be defined by its first-derivative's intervals w.r.t. attack-release onsets as follows:
\begin{equation}
\begin{split}
\begin{array}{cccc}
\textrm{Attack} &  \DERIVFLAT{W}{t} \geq 0 & 1 \leq t < T_{A} & \textrm{Attack-onset } T_A\\  
\textrm{Hold} & \DERIVFLAT{W}{t} = 0 & T_{A} \leq t < T_{R}&  \textrm{Release-onset } T_R\\  
\textrm{Release} & \DERIVFLAT{W}{t} \leq 0 & T_{R} \leq t \leq M & \textrm{Window-size } M. \\ 
\end{array}
\end{split}
\label{EQ:QPL:WIN_DYN}
\raisetag{1\baselineskip}
\end{equation}
The velocity, acceleration, and smoothness of $W(t)$ can be approximated via the following first-order forward, central, and squared central finite-differences of $\omega(t)$ respectively:
\begin{equation}
\begin{split}
 \DERIVFLAT{W}{t}  & \approx  \omega(t+1) - \omega(t) = \omega(t) * \nu_v(t),  \\
 \DERIVDERIVFLAT{W}{t}  &\approx  \omega(t+1) - 2\omega(t) + \omega(t-1) = \omega(t) * \nu_a(t), \\
\PR{\DERIVDERIVFLAT{W}{t}}^2  &  \approx  \omega(t) * \nu_{R}(t) * \omega(t), \quad \nu_{R}(t) = \nu_a(t) * \nu_a(t),
\end{split}
\label{EQ:QPL:FINITE_DIFF}
\raisetag{2\baselineskip}
\end{equation}
where $*$ is the discrete convolution operation, $\nu_{v}(t)$, $\nu_a(t)$, are the velocity, acceleration kernels respectively given by
\begin{equation}
\begin{split}
& \nu_v(t)   = \left \{  \begin{array}{rc}+1, & t = 1 \\ -1, & t = 0 \\ 0, & \textrm{otherwise} \end{array} \right . , \quad 
\nu_a(t)   = \left \{  \begin{array}{rc}+1, & t = \pm 1 \\ -2, & t = 0 \\ 0, & \textrm{otherwise} \end{array} \right . , 
\end{split}
\label{EQ:QPL:FINITE_DIFF_KER}
\raisetag{2.5\baselineskip}
\end{equation}
and $\nu_{R}(t)$ is the squared-acceleration smoothness kernel.
We maximize the overall smoothness of $W(t)$ with characteristic dynamics of \eqref{EQ:QPL:WIN_DYN} by minimizing the total squared-acceleration of \eqref{EQ:QPL:FINITE_DIFF} subject to causal COLA of \eqref{EQ:QPL:WINDOW} and finite-difference velocity constraints.

\begin{figure}[h]
  \centering
  {\includegraphics[width=0.9\linewidth]{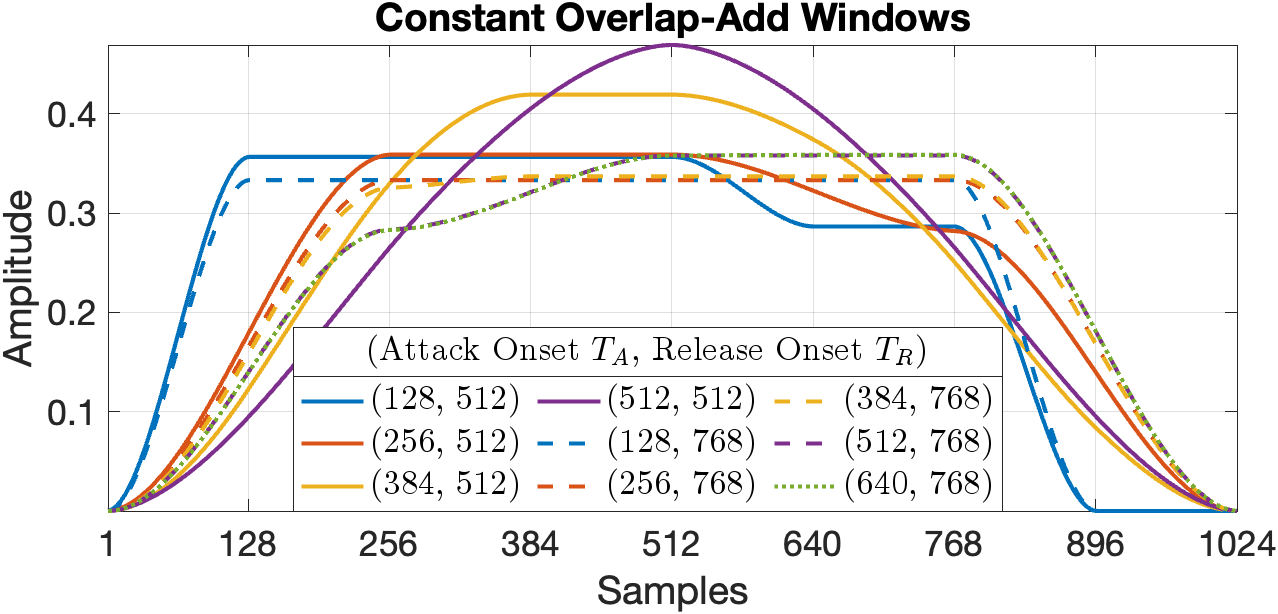}}
\caption{Solutions \eqref{EQ:QPL:WIN_OPT} vary across attack-onset $T_A$ and release-onset $T_R$  times, and are scale-invariant w.r.t the window/frame size ratio $M/F$.
}
\label{FIG:QPL:COLA}
\end{figure}

Let $\VEC{\omega} = \BK{\omega(1) ,\hdots, \omega(M)}^T \in \FIELD{R}^{M \times 1}$ be the unknown window samples, and the vectorized QP minimization be given by
\begin{equation}
\begin{split}
 \VEC{\omega}_*  &   = \ARGMIN{\VEC{\omega}} \sum_{t = 1}^{M} \PR{\omega(t) * \nu_a(t)}^2    = \VEC{\omega}^T \TOEPLITZBND{\nu_R}{0}{M} \VEC{\omega}, \quad \textrm{s.t.} \\
& \TOEPLITZBND{\nu_c}{0}{F} \VEC{\omega} = \VEC{1}, \quad
 \nu_c(t)  = \left \{ \begin{array}{lc}1, & t \equiv 0 \, (\bmod F) \\ 0, & \textrm{otherwise} \end{array}\right . ,  \quad 
 \VEC{\omega} \geq \VEC{0}, \\ 
& \TOEPLITZBND{\nu_v}{0}{T_A} \VEC{\omega}  \geq \VEC{0}, \quad
 \TOEPLITZBND{\nu_v}{T_A}{T_R} \VEC{\omega}  = \VEC{0}, \quad
 \TOEPLITZBND{\nu_v}{T_R}{M} \VEC{\omega}  \leq \VEC{0},  \\
\end{split}
\label{EQ:QPL:WIN_OPT}
\raisetag{1\baselineskip}
\end{equation}
where the listed constraints in \eqref{EQ:QPL:WIN_OPT} satisfy bounded and causal COLA in \eqref{EQ:QPL:WINDOW}, and attack-hold-release dynamics in \eqref{EQ:QPL:WIN_DYN}; a feasible solution can always be found for hold-sizes $T_R-T_A \leq \FLOOR{\frac{M}{F}}F$ as COLA rectangle windows can be shifted in time \cite{BROB_2012}. The operator  $\TOEPLITZBND{\nu}{a}{b} \in \FIELD{R}^{(b-a) \times M}$ in \eqref{EQ:QPL:WIN_OPT} defines a truncated Toeplitz matrix \cite{GRAY_TOEPLITZ_2006} with constant diagonal entries of the kernel function $\nu(t)$ given by
\begin{equation}
\begin{split}
T_{ij}  = \left \{ \begin{array}{cc}\nu(j - i - a), &   j-i \leq b \\0, & \textrm{otherwise} \end{array} \right . ,
\end{split}
\label{EQ:QPL:TOP}
\end{equation}
where $\nu(t)$ is shifted by $a$ and upper-bounded by $b$ in time.
The objective \eqref{EQ:QPL:WIN_OPT} is convex as the symmetric Toeplitz kernel matrix $\TOEPLITZBND{\nu_R}{0}{M}  = \PR{\TOEPLITZBND{\nu_a}{0}{M}}^2 + \VEC{e}_1 \VEC{e}_1^T  + \VEC{e}_M \VEC{e}_M^T$ can be decomposed into the sum of positive semi-definite (PSD) products of the symmetric Toeplitz matrix of $\nu_a(t)$ in \eqref{EQ:QPL:FINITE_DIFF} with itself and non-negative outer products.
%
Furthermore, $\TOEPLITZBND{\nu_R}{0}{M}$ is consistent with the $0$-value Dirichlet boundary conditions of \eqref{EQ:QPL:WINDOW}, causing window tails to taper in the solutions to \eqref{EQ:QPL:WIN_OPT}, as shown in Fig. \ref{FIG:QPL:COLA} $(M=1024, \, F = 256)$; several regularities for attack-release onset tuples $(T_A, T_R)$ are noted.

Symmetric onsets times, $T_A = M - T_R$ (solid-purple, dashed-red), yield symmetric windows with shapes following convolved rectangle-cosine  constructions in \cite{BROB_2012}. Late attack-onsets $T_A \geq M/2$ (dashed-purple, dotted-green) present identical asymmetric windows. Early attack-onsets $T_A < F$ (solid-blue) exhibit piece-wise flat regions. The remaining cases (red, yellow) have a single flat section in the hold-interval $(T_A, T_R)$. Lastly, solutions preserve their shapes for constant $M/F$, and suggest a continuous and closed-form expression.


\section{Channel-Mixture Distortion Objective}
\label{SEC:OBJ}

We define channel-mixture distortion as the weighted summation of channel gain reductions in decibels (dB), which can be expressed as the logarithmic product $g(\VEC{x})$ of exponentiated gain variables $x_n^{w_n}$:   \vspace{-1.5mm}
\begin{equation}
\begin{split}
\sum_{n=1}^N   w_n  20  \log_{10} x_n = 20 \log_{10}  g(\VEC{x}), \quad
  g(\VEC{x})   =   \prod_{n=1}^N x^{w_n}_n,
\end{split}
\label{EQ:OBJ:LOUD}
\end{equation}
where $w_n \in \FIELD{R}_{>0}$ is the positive attenuation rate parameterizing dB reduction of channel $Y(t, n)$ in \eqref{EQ:QPL:ENVELOPE} per dB reduction to $x_n$. Larger $w_n$ penalizes smaller $x_n$ and minimize distortion to select channels (e.g. center). The maximizer of the summation in \eqref{EQ:OBJ:LOUD} is equal to that of the product of gains as the logarithm is both monotonic and non-positive. For QP objective \eqref{EQ:QPL:OBJ}, the quadratic approximation of $g(\VEC{x})$ follows the Taylor series expansion $h(\VEC{x})$  at an expansion center $\VEC{a} \in \FIELD{R}^{N \times 1}$:
\begin{equation}
\begin{split}
h(\VEC{x})    =  g(\VEC{a}) +  \PR{\VEC{x} - \VEC{a}}^T \nabla g(\VEC{a})   
 + \frac{1}{2} \PR{\VEC{x} - \VEC{a}}^T \MAT{H}(\VEC{a}) \PR{\VEC{x} - \VEC{a}}, \\
\end{split}
\label{EQ:OBJ:LOUD_APPX}
\raisetag{1.75\baselineskip}
\end{equation}
where the gradient $\nabla g(\VEC{x})  \in \FIELD{R}^{N \times 1}$ is the vector of partial derivatives
\begin{equation}
\begin{split}
\nabla g(\VEC{x}) = \BK{\PDERIV{g}{x_1}, \hdots, \PDERIV{g}{x_N} }^T, \quad 
\PDERIV{g}{x_i}  = w_i x^{w_i \minus 1}_i\prod_{\substack{n=1, \,  n \neq i}}^N x^{w_n}_n, 
\end{split}
\label{EQ:OBJ:LOUD_APPX_GRAD}
\raisetag{0.666\baselineskip}
\end{equation}
and the Hessian $\MAT{H}(\VEC{x}) \in \FIELD{R}^{N \times 1}$ is the matrix of second-order partial derivatives $H_{ij} = \PDERIVTWO{g}{x_i}{x_j}$ with entries given by
\begin{equation}
\begin{split}
H_{ij}  = 
 \left \{ \begin{array}{lc}  w_i (w_i - 1) x^{w_i \minus 2}_i \prod_{\substack{n=1 \\ n \neq i}}^N x^{w_n}_n , & i = j  \\[2mm] w_i w_j x_i^{w_i \minus 1} x_j^{w_j \minus 1} \prod_{\substack{n=1 \\ n \neq i, j}}^N x^{w_n}_n, & i \neq j  \end{array}   \right . .
\end{split}
\label{EQ:OBJ:LOUD_APPX_HESS}
\end{equation}
The expansion center at unity $\VEC{a} = \VEC{1}$  in  \eqref{EQ:OBJ:LOUD_APPX} incurs no distortion to \eqref{EQ:OBJ:LOUD}, and has the following gradient \eqref{EQ:OBJ:LOUD_APPX_GRAD} and Hessian \eqref{EQ:OBJ:LOUD_APPX_HESS}: 
\begin{equation}
\begin{split}
 g(\VEC{1}) = 1, \, \, \, \,  \nabla g(\VEC{1}) = \VEC{w}, \, \, \, \, \MAT{H}(\VEC{1}) =  \VEC{w} \VEC{w}^T - \DIAGF{\VEC{w}},
\end{split}
\label{EQ:OBJ:LOUD_APPX_QUAD}
\end{equation}
where $\VEC{w}  = \BK{w_1, \hdots, w_N}^T \in \FIELD{R}^{N \times 1}$ is the vector of attenuation rates, and  $\DIAGF{\VEC{w}} = \sum_{n=1}^N w_n \VEC{e}_n$ the diagonal operator. Rewriting \eqref{EQ:OBJ:LOUD_APPX}, \eqref{EQ:OBJ:LOUD_APPX_QUAD} in the standard form \eqref{EQ:QPL:OBJ} for QP minimization gives
\begin{equation}
\begin{split}
f(\VEC{x}) & = 1 - h(\VEC{x}), \quad
\MAT{Q}  =  -\MAT{H}(\VEC{1}) = \DIAGF{\VEC{w}} - \VEC{w} \VEC{w}^T, \\
 \VEC{c} & = -\PR{\VEC{w} + \MAT{Q} \VEC{1}} = (\VEC{w}^T \VEC{1} - 2) \VEC{w} , \quad 
 d = \frac{1}{2} \VEC{1}^T \MAT{Q} \VEC{1} + \VEC{w}^T \VEC{1}, \\
\end{split}
\label{EQ:OBJ:LOUD_OBJ}
\raisetag{2.0\baselineskip}
\end{equation}
where it is desirable to further constrain the critical point $\VEC{x}_{*}$ of $f(\VEC{x})$ to $\VEC{x}_{*} \geq \VEC{1}$ such that the minimizer of \eqref{EQ:QPL:CONSTR} lies on a constraint that is near unity. Setting the gradient $ \nabla f(\VEC{x})  = \MAT{Q} \PR{\VEC{x}_{*} - \VEC{1}} - \VEC{w}$ to zero, and computing the Sherman-Morrison \cite{SHERMAN_MORRISON_1950} matrix-inverse $\MAT{Q}^{\minus 1} = \DIAGF{\VEC{w}}^{\minus 1} + \frac{\VEC{1} \VEC{1}^T}{ 1 - \VEC{1}^T \VEC{w}}$, yields the critical point given by 
\vspace{-1.5mm}
\begin{equation}
\begin{split}
 \VEC{x}_{*} & = \MAT{Q}^{\minus 1}\VEC{w} + \VEC{1} = \PR{2 + q} \VEC{1}, \quad 
  q = \VEC{1}^T \VEC{w} \PR{1 - \VEC{1}^T\VEC{w}}^{-1}, 
\end{split}
\label{EQ:OBJ:LOUD_OBJ_CRIT}
\end{equation}
where  $\VEC{x}_{*}$ is a scaled vector of unity. Therefore, constraining  $\VEC{x}_{*}$ to lie outside of unity bounds the sum of rates $w_n$ below unity:
\begin{equation}
\begin{split}
    \VEC{x}_{*} & \geq \VEC{1} \,     \Leftrightarrow  \, q \geq -1 \,     \Leftrightarrow \,  \sum_{n=1}^N w_n  \leq 1, \quad 
    w_n > 0.
\end{split}
\label{EQ:OBJ:LOUD_OBJ_RATE_BND}
\end{equation}

A second desirable property is a PSD $\MAT{Q} \succeq 0$ or negative semi-definite $\MAT{H}(\VEC{1}) \preceq 0$ to ensure that $f(\VEC{x})$ is convex in \eqref{EQ:QPL:CONSTR} and solvable in polynomial time \cite{KOZLOV_1980}. We show that the summation bound $\sum_{n=1}^N w_n \leq 1$ in  \eqref{EQ:OBJ:LOUD_OBJ_RATE_BND} is necessary and sufficient for PSD $\MAT{Q}$. Let the eigenvalues $\lambda  \in \FIELD{R}$ of symmetric $\MAT{Q}$ be the zeros of secular equation $S(\lambda)$ of a rank-1 update to a diagonal matrix \cite{GOLUB_1973_EIG}:
\begin{equation}
\begin{split}
S(\lambda) & = 1 - \sum_{n=1}^N \frac{w_n^2}{w_n - \lambda} 
= 1 - \sum_{n=1}^N \frac{\PR{w_n^2 - \lambda^2} + \lambda^2 }{w_n - \lambda} \\
& = 1 - \sum_{n=1}^N w_n - \lambda \sum_{n=1}^N \frac{ w_n}{w_n - \lambda}. \quad \textrm{Diff. of Squares} 
\raisetag{1\baselineskip}
\end{split}
\label{EQ:OBJ:SECULAR}
\end{equation}
Setting \eqref{EQ:OBJ:SECULAR} to zero and applying bounds \eqref{EQ:OBJ:LOUD_OBJ_RATE_BND} gives the necessary conditions that constrain the eigenvalues as follows:
\begin{equation}
\begin{split}
\sum_{n=1}^N w_n = 1 -  \lambda \sum_{n=1}^N \frac{  w_n}{w_n - \lambda} \leq 1  \,  \Rightarrow \,  \lambda \geq 0,
\end{split}
\label{EQ:OBJ:PSD_NEC}
\end{equation}
whereby $\lambda$ must be non-negative via proof by contrapositive as $\lambda < 0 \, \Rightarrow \, \sum_{n=1}^N  \frac{  w_n}{w_n - \lambda} > 0 \, \Rightarrow \, 1- \lambda \sum_{n=1}^N  \frac{  w_n}{w_n - \lambda} > 1$.
For the sufficient condition, let rates $0 < w_1 \leq w_2 \leq \hdots w_N$ and eigenvalues $\lambda_1 \leq \lambda_2 \leq \hdots \lambda_N$ of $\MAT{Q}$ be sorted in ascending order. We prove $\lambda_1 \geq 0 \Rightarrow \sum_{n=1}^N w_n \leq 1 $ via the contrapositive  $\sum_{n=1}^N w_n > 1 \Rightarrow \lambda_1 < 0 < \lambda_2$ and examine $S(\lambda)$ in \eqref{EQ:OBJ:SECULAR} as follows:
\vspace{-3.5mm}
\begin{equation}
\begin{split}
&  S(-\infty) = 1,  \quad S(0) = 1- \sum_{n=1}^N w_n < 0, \\[-5pt]
\lim_{\lambda \rightarrow w_n^{\pm}}S(\lambda) & = \pm \infty, \, \quad
 \PDERIV{S(\lambda)}{\lambda} = -\sum_{n=1}^N \frac{w_n^2}{\PR{w_n - \lambda}^2} < 0,
\end{split}
\label{EQ:OBJ:PSD_SUFF}
\end{equation}
where the partial derivative $\PDERIVFLAT{S(\lambda)}{\lambda}$ is negative. The intermediate value theorem therefore guarantees the existence of the two smallest roots in the intervals $\lambda_1 < 0$ and $w_1 < \lambda_2 < w_2$ respectively. Thus in practical implementations, we normalize $w_n$ to sum to or below unity.

\section{Variable and Constraint Reduction}
\label{SEC:VCR}

QP solvers exhibit quadratic to cubic compute costs w.r.t. the number of constraints and variables \cite{NEMIROVSKI_INT_2004}, \cite{OSQP_STELLATO_2020} in \eqref{EQ:QPL:QPFRAME}, which is prohibitive for real-time processing within an audio frame's period in practical applications. Consider the case of multi-band, multi-content channel inputs with multiple output-mixer terminals. Let the channel gains be the column matrix $\MAT{X} \in \FIELD{R}^{N_B \times N_C}$ of $N_B$ multi-band gains per $N_C$ content channels, and the sample data be the tensor $\MAT{S} \in \FIELD{R}^{\bar{F} \times N_B \times N_C \times N_M}$ of $\bar{F}$ samples per frame and look-ahead, $N_B$ bands, $N_C$ content channels, summed into $N_M$ mixers. The constraint set is given by 
\begin{equation}
\begin{split}
 -  \tau_m \leq  \sum_{j=1}^{N_B}  \sum_{k=1}^{N_C}   S(i, j, k, m) \, X_{jk}     \leq \tau_m, \quad 
 0  \leq X_{jk} \leq  1, \\ 
\end{split}
\label{EQ:VCR:CONSTR}
\end{equation}
where $1 \leq i \leq \bar{F}$, $1 \leq m \leq N_M$, and the variables $\VEC{x} = \VECT{\MAT{X}} \in \FIELD{R}^{N \times 1}$  can be vectorized by stacking the columns of matrix $\MAT{X}$ for consistency with \ref{EQ:QPL:QPFRAME}. Both the number of variables $N = N_B N_C$, and sample constraints $M = 2 \bar{F} N_M$ are multiplicative and large. We give two methods for reducing the number of variables and constraints.

\textbf{Pre-mixing Variable Reduction:}
We can consider pre-mixing subsets of input channels as an approximation to the QP in \ref{EQ:QPL:QPFRAME} that constrains the variables $\VEC{x}$ to a linear transformation $\VEC{x} = \MAT{P} \VEC{y}$, $\MAT{P} \in \FIELD{R}^{N \times N_P}$  of fewer  $N_P < N$ variables in vector $\VEC{y} \in \FIELD{R}^{N_P \times 1}$:
\begin{equation}
\begin{split}
 \VEC{y}_{*}  = \ARGMIN{\VEC{y}}  f(\VEC{y})   = \frac{1}{2} \VEC{y}^T \MAT{P}^T \MAT{Q}  \MAT{P} \VEC{y} + \VEC{c}^T \MAT{P} + d, \\
\textrm{s.t.   } -\VEC{\tau}_m  \leq \MAT{S}_m \MAT{P} \VEC{y}  \leq \VEC{\tau}_m, \quad
\VEC{0} \leq \MAT{P} \VEC{y} \leq \VEC{u}, \quad
\end{split}
\label{EQ:VCR:LOWRANK}
\end{equation}
whereby $\MAT{P}^T \MAT{Q} \MAT{P} \succeq 0$ preserves convexity, channel matrix $\MAT{S}_m \in \FIELD{R}^{\bar{F} \times N}$ reshapes $S(:,:,:, m)$ in \eqref{EQ:VCR:CONSTR}, $\VEC{\tau}_m = \tau_m \VEC{1} \in \FIELD{R}^{\bar{F} \times 1}$ specifies mixer-dependent thresholds,  and  $\MAT{S}_m \MAT{P}$ downmixes channel samples. The following pre-mixer matrices $\MAT{P}$ are lossless if $\MAT{P} \VEC{1} = \VEC{1}$, uncouples variables in $\VEC{y}$ if each row contains exactly one non-zero entry, and induces finite upper-bounds on $\VEC{y}$ if all entrants are non-negative:
\begin{equation}
\begin{split}
\left . \arraycolsep=1.7pt\def\arraystretch{0.0}
  \begin{array}{ccccc}
  \textbf{Pre-mixer} &  \textbf{Definition } \MAT{P} &  N_P & \textbf{Uncoupled} \\  [0.5mm]   \hline  \noalign{\vskip 1mm}  
  \textrm{Single channel} & \VEC{1}_{N} & 1  & \textrm{True} \\
  \textrm{Multi-band} &  \MAT{I}_{N_C} \otimes \VEC{1}_{N_B} & N_C & \textrm{True} \\
    \textrm{Multi-content} &  \VEC{1}_{N_C}\otimes \MAT{I}_{N_B} & N_B  & \textrm{True} \\  [0.5mm]
  \textrm{Concatenation} & \begin{array}{c}[\alpha \VEC{1}_{N_C}\otimes \MAT{I}_{N_B}, \\ [0.25mm] \MAT{I}_{N_C} \otimes (1 - \alpha)  \VEC{1}_{N_B}] \end{array} & N_B + N_C  & \textrm{False} \\
  \end{array}
  \right .
\end{split}
\label{EQ:VCR:PREMIX}
\end{equation}
where  $\otimes$ is the Kronecker product operator. The single, multi-band, and multi-content pre-mixers sums all, banded, and content channels respectively. The concatenation pre-mixer contains multi-band, and multi-content weighted downmixes with lossless  $0 < \alpha < 1$, and box-constraints $\VEC{0} \leq \VEC{y} \leq \BK{\alpha^{\minus 1} \VEC{1}_{N_B} ; (1- \alpha)^{ \minus 1} \VEC{1}_{N_C}}$. We omit downmix matrices that are both lossy and coupled (cross-format AC-3 \cite{ATSC_2012_DOWNMIX}), but can be used to preserve left-right content balance in practice.

\begin{figure}[ht]
  \centering
  {\includegraphics[width=0.875\linewidth]{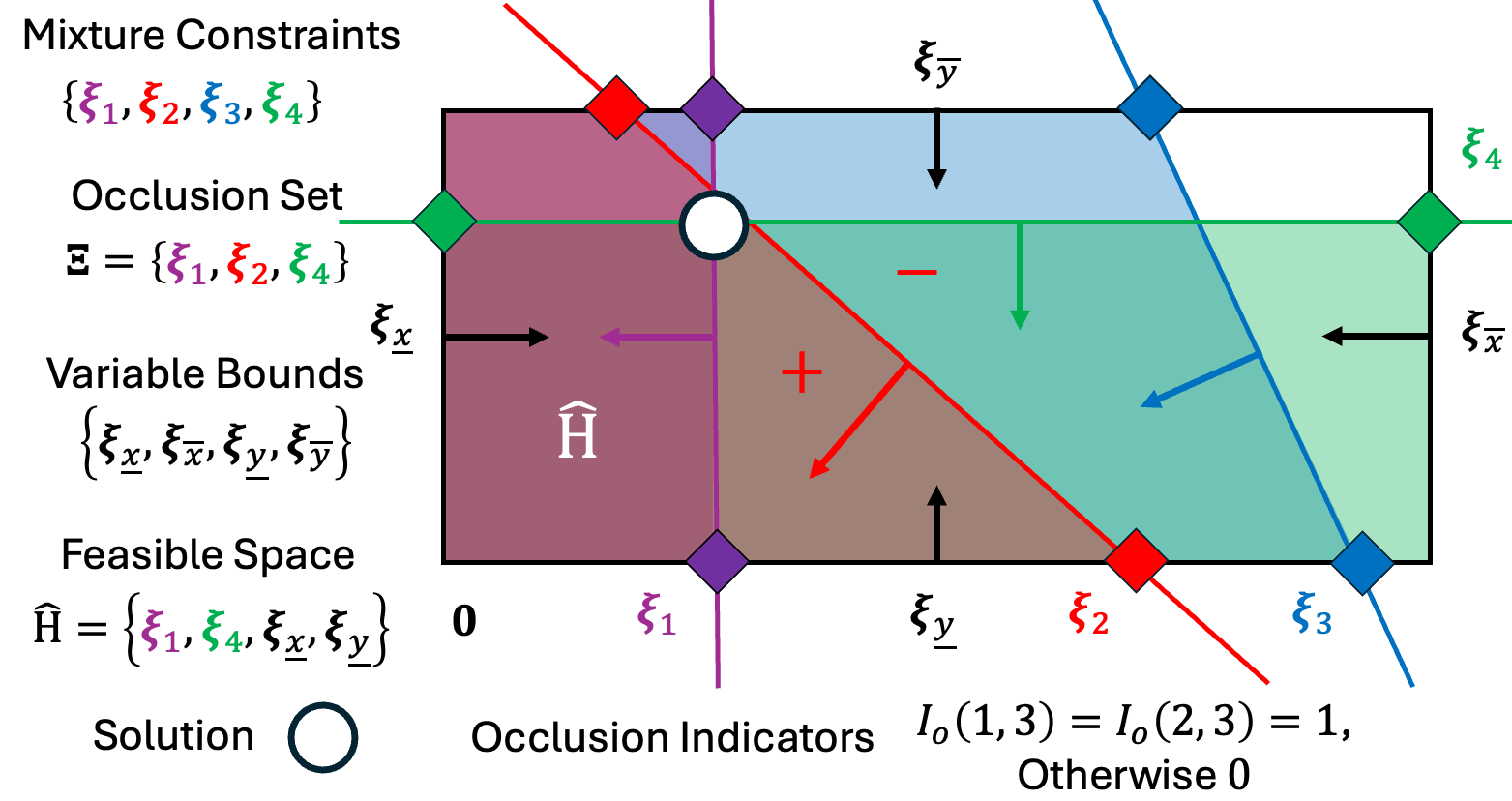}}
\caption{The sample constraint set's feasible space $\hat{\FIELD{H}}$ is contained in the non-occluded set of constraints $\MAT{\Xi}$ in \eqref{EQ:VCR:MIN_OCC_SET} which excludes $\xi_3$. }
\label{FIG:VCR:CONSTRAINTS}
\end{figure}

\textbf{Occlusion Culling Constraint Reduction:} We can efficiently identify and remove a class of mixture limit constraints in \eqref{EQ:QPL:CONSTR_SAMP} that do not support the QPs feasible space.
Let $\MAT{V}^{\CR{m}}$ be the set of extreme vertices in $\FIELD{R}^{N}$ of the convex hull $\FIELD{H}^{\CR{m}}$ defined by the following half-space intersection of the signed $\ABS{m}^{th}$ mixture constraint $s_{m}(\VEC{x}) =  \SGN{m} \sum_{n=1}^N S_{\ABS{m} n} x_n \leq \tau$  and the box-constraints in \eqref{EQ:QPL:QPFRAME}:
\begin{equation}
\begin{split}
\FIELD{H}^{\CR{m}} =  \CR{\VEC{x} \in \FIELD{R}^N :  \, s_{m}(\VEC{x}) \leq \tau \,  \wedge   \,   \VEC{0} \leq  \VEC{x} \leq \VEC{u} },
\end{split}
\label{EQ:VCR:HULL}
\end{equation}
where $\SGN{}$ is signum, $\underline{\MAT{V}}^{\CR{m}} \subset \MAT{V}^{\CR{m}}$ is the subset of vertices that also lie in equality of the  $m^{th}$ constraint $s_{m}(\VEC{x}) =  \tau$, and has cardinality $| {\underline{\MAT{V}}^{\CR{m}} }  | \leq 2^{N \minus 1}$  bounded by maximum number of edges $2^{N \minus 1}N$ in the $N$-dimensional hyper-cube. A convex hull  $\FIELD{H}^{\CR{i}}$ is fully contained in convex hull $\FIELD{H}^{\CR{j}}$ iff the $i^{th}$ mixture constraint \textit{occludes} the $j^{th}$ mixture constraint via the following indicator function:
\begin{equation}
\begin{split}
I_o(i, j) = \left \{ \begin{array}{cc} 1, & 
s_i(\VEC{x}) > \tau, \quad \forall \VEC{x} \in \underline{\MAT{V}}^{\CR{j}} \\
0, & \textrm{otherwise}
\end{array}
 \right . ,  \quad 
\end{split}
\label{EQ:VCR:OCC_IND}
\end{equation}
whereby all vertices in $\underline{\MAT{V}}^{\CR{j}}$ lie on the negative half-space of the $i^{th}$ constraint. The $j^{th}$ constraint $s_j(\VEC{x}) \leq \tau$ can therefore be removed from constraint set $\VEC{\xi}$ in \eqref{EQ:QPL:QPFRAME} as it does not support the intersection of convex hulls $\hat{\FIELD{H}} = \cap_{i} \FIELD{H}^{\CR{i}}$ that defines the feasible space in Fig. \ref{FIG:VCR:CONSTRAINTS}. We compare the costs of finding $\hat{\FIELD{H}}$ to solving the QP given $M$ number of constraints and $\bar{M} \leq M$ supports of $\hat{\FIELD{H}}$ in $N$ variables, where $M \gg N$.

Let the cost of finding the support constraints of $\hat{\FIELD{H}}$ via dual space methods \cite{BARBER_QHULL_1996} be $O(M \bar{M}^{\FLOOR{N/2}} / (\bar{M} \FLOOR{N/2}!) )$, which exceeds that of solving the QP via interior-point methods \cite{NEMIROVSKI_INT_2004} in $O(M^{3/2} N^2)$ or first-order methods \cite{OSQP_STELLATO_2020} in $O((M+N)^3)$. It is therefore efficient to solve the QP \eqref{EQ:QPL:QPFRAME} over a small super-set of support constraints, initially determined by pre-processing \cite{GOULD_PRESOLVE_2004} in $O(MN)$, before applying further constraint reduction in sub-quadratic time. Consider the following set $\MAT{\Xi}$ of mixture constraints not occluded by any other constraints:
\begin{equation}
\begin{split}
\MAT{\Xi} = \CR{ s_j(\VEC{x}) \leq \tau: \,  I_o(i, j) = 0, \, \forall \, i \neq j  }, \quad 
\hat{M} = \ABS{\MAT{\Xi}},
\end{split}
\label{EQ:VCR:MIN_OCC_SET}
\end{equation}
which contains the supports of $\hat{\FIELD{H}}$. The costs of computing  $\underline{\MAT{V}}^{\CR{j}}$ and evaluating $I_o(i, j)$ in \eqref{EQ:VCR:OCC_IND} are $O(2^{N} N^2)$ and  $O(2^{N} N)$ respectively. We can determine $\MAT{\Xi}$ in $O(2^{N} (N^2 + N \hat{M} + \log M) M )$ time via sorted constraints defined in the following propositions:
\begin{equation}
\begin{split}
 I_o(i, j)   = 1 \quad \Rightarrow  \quad 
   \min_{\VEC{x} \in \underline{\MAT{V}}^{\CR{i}} } \NORM{\VEC{x}}  \, \,    < \min_{\VEC{x} \in \underline{\MAT{V}}^{\CR{j}} }  \NORM{\VEC{x}}, 
\end{split}
\label{EQ:VCR:MIN_DIST}
\end{equation}
whereby the vertex in $\underline{\MAT{V}}^{\CR{i}} $ closest to the origin is closer than that of $\underline{\MAT{V}}^{\CR{j}}$ if constraint $i$ occludes constraint $j$ following $\FIELD{H}^{\CR{i}} \subset \FIELD{H}^{\CR{j}}$. The contrapositive of \eqref{EQ:VCR:MIN_DIST} is therefore true and expressed by
\begin{equation}
\begin{split}
       \min_{\VEC{x} \in \underline{\MAT{V}}^{\CR{i}} } \NORM{\VEC{x}}  \, \,    \geq \min_{\VEC{x} \in \underline{\MAT{V}}^{\CR{j}} } \NORM{\VEC{x}} \quad
      \Rightarrow  \quad  I_o(i, j) = 0,
\end{split}
\label{EQ:VCR:CONT}
\end{equation}
where if the vertex in $\underline{\MAT{V}}^{\CR{i}} $ closest to the origin is further than or equidistant to that of $\underline{\MAT{V}}^{\CR{j}}$, then constraint $i$ does not occlude constraint $j$. We can therefore sort the $M$ constraints in ascending order by the min-norm vertex $\min_{\VEC{x} \in \underline{\MAT{V}} } \NORM{\VEC{x}}$ in $O(2^{N} M \log M)$ time, add the first constraint to $\MAT{\Xi}$ for the base case, and add subsequent sorted constraints $j$ to $\MAT{\Xi}$ if all constraints $i$ in  $\MAT{\Xi}$ do not occlude constraint $j$  $\PR{I_0(i, j) = 0, \, \forall s_i(\VEC{x}) \leq \tau \in \MAT{\Xi}}$ in $O(2^{N} N  \hat{M} M )$ time.

\section{Experiments}
\label{SEC:EXP}

\begin{figure}[ht]
  \centering
  {\includegraphics[width=0.835\linewidth]{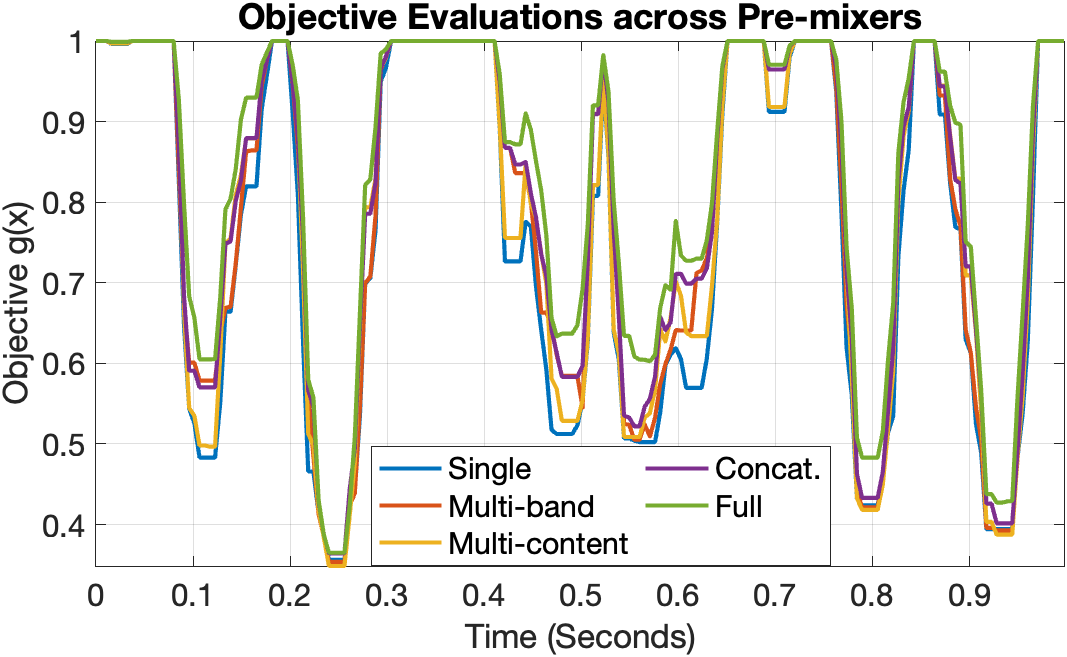}}
\caption{Distortion objective $g(x)$ in \eqref{EQ:OBJ:LOUD} for \eqref{EQ:VCR:PREMIX} pre-mixers converges to the full mixer-limiter as the number of channels increases.}
\label{FIG:EXP:PREMIX}
\end{figure}

We evaluate the distortion objective $g(\VEC{x})$ in \eqref{EQ:OBJ:LOUD} in terms of the QP objective $f(\VEC{x})$ in \eqref{EQ:QPL:QPFRAME} across pre-mixers in \eqref{EQ:VCR:PREMIX} of decreasing number of channels compared to the full mixer-limiter. The multi-band and mult-content channel tensor $\MAT{S}$ in \eqref{EQ:VCR:CONSTR} is populated by  amplitude modulated signals  $S(t, j, k)  =  \sin \PR{2 \pi a_j t  } \sin \PR{2 \pi (b_k t + \phi_{jk}) }$,
where $\VEC{a} = \BK{101, 443, 1627}$ and $\VEC{b} = \BK{2, 5, 11}$ are lists of frequencies (Hz) for  the carrier band and message content respectively, and $\phi_{jk} =  \frac{(k-1) N_B + j}{N_B N_C}$ the latter's phase-offset. We plot the time-evolution of $g(\VEC{x})$ over a $1$-second duration in Fig. \ref{FIG:EXP:PREMIX}, and show the distortion improvement gap with the full mixer-limiter decreases for summative $N_B + N_C$ number of channels. The mean and standard deviation of $f(\VEC{x})$ across pre-mixers validate the trend: Single channel limiter $0.23 \pm 0.23$, multi-band and multi-content channel limiters $0.2 \pm 0.21$, concatenation ($\alpha = 1/2$) $0.19 \pm 0.2$, and full $0.16 \pm 0.18$.

We evaluate constraint reduction performance by simulating a multi-band mixer-limiter in \eqref{EQ:QPL:QPFRAME} of frame-size $F = 256$, look-ahead $L = 768$, mixture threshold  $\VEC{\tau} = \VEC{1}$, and upper bound  $\VEC{u} = \VEC{1}$. The input signals consist of $N$ full-scale sine tones at $\BK{101,  443,  1627,  4153,  8747,  15733}$ Hz ($1$-second duration at $48$ kHz sample rate). Pre-processing \cite{GOULD_PRESOLVE_2004} implied-bounds and tightening $\VEC{x}$ methods reduce the number of mixture-limit constraints (originally $2048$). The non-occluded sets $\MAT{\Xi}$ \eqref{EQ:VCR:MIN_OCC_SET} computed with or without pre-processing are equivalent. We compare its cardinality $\hat{M}$ to that of the convex hull's supports in Table \ref{TAB:EXP:MULTI_CONSTR} across frames for mixer-limiter instances of increasing number of channels, containing the first $N$ tones. The ratio of constraint to convex-support set sizes reduces in range from $(8.07, 51.24)$ (pre-processing) to $(1.37, 1.88)$ (non-occluding set), thereby approaching the cardinality's lower-limit.


\begin{table}[h]
  \vspace{1.75mm}
\centering
\caption{Mixer-limiter Number of Constraints (Mean $\pm$ Std. Dev)}
\label{TAB:EXP:MULTI_CONSTR}
\begingroup
\setlength{\tabcolsep}{4pt}
  \begin{tabular}{cllll}
 \textbf{$N$} &  \textbf{Implied-bounds} & \textbf{Tightening} & \textbf{Non-occluded} & \textbf{Convex} \\   [0.35mm]   \hline  \noalign{\vskip 1mm}      
 2 &  $ 384.7 \pm 52 $ & $ 374.1 \pm 50.5 $ & $ 10 \pm 4.3 $ & $ 7.3 \pm 2.6 $ \\
   3 &  $ 805.8 \pm 103.7 $ & $ 799 \pm 102.8 $ & $ 41.8 \pm 14.7 $ & $ 25.9 \pm 7.5 $ \\
   4 &  $ 1167 \pm 149.4 $ & $ 1164 \pm 149 $ & $ 99.1 \pm 22.9 $ & $ 58.5 \pm 14 $ \\
   5 &  $ 1442 \pm 184.5 $ & $ 1441 \pm 184.3 $ & $ 226.3 \pm 64 $ & $ 130.1 \pm 35.7 $ \\
   6 &  $ 1636 \pm 209.3 $ & $ 1636 \pm 209.2 $ & $ 381.5 \pm 78.6 $ & $ 202.8 \pm 41.8 $
  \end{tabular}
  \endgroup
    \vspace{-3.5mm}
\end{table} 

\section{Conclusions}
We presented a coupled multichannel mixer-limiter design for adaptive channel-headroom allocation and loudspeaker protection. A minimum distortion objective was approximated and optimized via a QP formulation over a frame-based processor. Limiter envelopes satisfying the QP constraints were constructed from dynamics-constrained COLA window formulations. Pre-mixing and occlusion-set variable-constraint reduction methods decreased the QP problem size and achieved comparable performance with the full mixer-limiter in experiments.

\clearpage

\section{Acknowledgment}
This work is funded by Amazon.com incorporated.

\section{Compliance with Ethical Standards}
This is an audio signal-processing study for which no ethical approval was required.

\bibliographystyle{IEEEbib}
\bibliography{IEEEabrv,refs}


\clearpage


\end{document}